# Technical Insights on Blockchain's Role in Financial Systems

1st Ishan Patwardhan
*Department of Computer Engineering*
*COEP Technological University*
Pune, India
patwardhanip20.comp@coeptech.ac.in

2nd Dr Sunil Mane
*Department of Computer Engineering*
*COEP Technological University*
Pune, India
sunilbmane.comp@coeptech.ac.in

3rd Nidhi Patel
*Department of Computer Engineering*
*COEP Technological University*
Pune, India
nidhip20.comp@coeptech.ac.in

*Abstract*—This research provides a critical analysis regarding the way blockchain is being implemented in the financial industry, highlighting its vital role in promoting green finance, guaranteeing compliance with regulations, improving supply chain finance, boosting decentralized finance (DeFi), and strengthening the Internet of Things (IoT). It discusses how blockchain's inherent attributes could significantly boost transparency, operational efficiency, and security across these domains while also addressing the pressing challenges of scalability, system integration, and the evolving regulatory landscape.

## I. INTRODUCTION

A new era of innovation and disruption has been brought about by blockchain technology, with the banking industry being one of the most affected. Blockchain's decentralized, transparent, and immutable nature offers unprecedented opportunities for enhancing efficiency, reducing costs, and improving security in financial systems [1] [2]. While initially conceived to support cryptocurrencies like Bitcoin, the technology's scope has expanded to include a multitude of financial applications, from green finance and supply chain finance to decentralized finance (DeFi) and regulatory compliance [3] [4].

One of the most intriguing applications of blockchain in finance is its potential to revolutionize green finance. Blockchain technology has the potential to streamline access to alternative avenues of finance and investment, notably by attracting funds from private investors. It functions within decentralized frameworks, circumventing conventional financial intermediaries [3]. This not only decreases costs but also enhances transparency, thereby reducing the risk of "greenwashing" [3].

In addition to green finance, blockchain technology is making significant inroads into the realm of Internet of Things (IoT) in financial applications. Hyperledger Fabric, for instance, has been integrated with IoT devices to establish a root of trust and demonstrate access control, thereby enhancing security [5]. This integration is particularly crucial given the increasing role of IoT in financial services for data collection and automation.

Moreover, blockchain's impact is not limited to developed economies; it is also making its mark in emerging markets. For example, in Africa, blockchain technology is being explored for its potential to either reinforce or disrupt existing financial and security infrastructures [6]. However, this also raises questions about the technology's role in perpetuating or alleviating existing socio-economic inequities [6].

Despite its promise, blockchain technology faces challenges that could hinder its widespread adoption in finance. These include issues related to scalability, security, and regulatory compliance [2]. Understanding these challenges is crucial for both academics and practitioners to unlock the full potential of blockchain in finance.

This research endeavors to present a thorough review for the current state of blockchain applications in finance, exploring its various uses, benefits, and challenges. Through this review, we seek to offer insights into how blockchain can shape the future of finance, from enhancing green investments to improving regulatory compliance and beyond.

## II. LITERATURE REVIEW

This research includes a web scraping technique in addition to a thorough literature search carried out using scholarly databases such as PubMed, IEEE Xplore, and Google Scholar. Using the arXiv.org database, we created a unique web scraper to collect more publications and papers. Blockchain, finance, green finance, supply chain finance, regulatory compliance, Decentralised Finance, Internet of Things, and emerging markets are just a few of the terms that this scraper was designed to look for in articles. Our research was able to cover a wider range of topics because of this method's ability to gather more recent and pertinent material.

### A. Blockchain and Green Finance

The emerging integration of blockchain in green finance is pivotal for sustainability, aiding in the reduction of carbon emissions and supporting renewable energy initiatives. Research in this area reveals how green finance significantly reduces agricultural carbon emissions by influencing the use of chemical fertilizers [7]. Blockchain's role is crucial in this context, as it ensures transparency and accountability in green finance, specifically in its allocation and use for sustainable agricultural methods. This highlights blockchain's potential

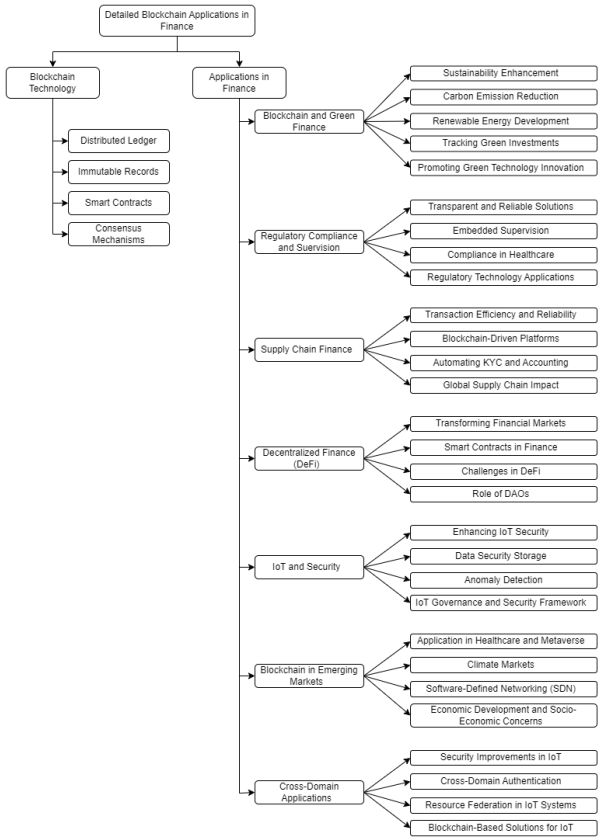

Fig. 1. Topological diagram of Blockchain and its applications

in both monitoring and verifying green investments, ensuring their effective implementation.

Employing fuzzy-set Qualitative Comparative Analysis (fsQCA) in the investigation of green finance's impact on carbon emissions, the study seeks to scrutinize the intricate interactions and determinants influencing environmental deterioration [8]. The integration of blockchain in this field significantly enhances data accuracy and reliability, offering valuable insights for crafting effective green finance policies.

Separately, research emphasizes how green finance can effectively adjust the effects of diverse green regulations on green technology innovation. The decentralized and unchangeable nature of blockchain technology strengthens role of green finance in encouraging the development of green technology, ensuring adherence to diverse environmental standards [9].

Delving into the domain of sustainable inter-business e-commerce and the financial management of supply chains, an in-depth analysis unveiled the crucial significance of blockchain technology in enhancing operational efficacy, notably within logistics and digital record-keeping [10]. The inherent attributes of blockchain, including its capacity for ensuring data integrity, transparency, and security, are identified as key enhancers of reliability and efficacy in green finance applications within sustainable supply chain management frameworks.

Simultaneously, an in-depth examination of the impact of green finance on mitigating climate variability emphasized the critical importance of investing in renewable energy infrastructure. [11]. Blockchain technology emerges as a critical facilitator in this context, streamlining secure and efficient financial transactions. This technological integration is instrumental in bolstering investments in renewable energy initiatives, thereby contributing significantly to the diminishment of carbon emissions and the broader endeavor to mitigate the repercussions of climate change.

The study emphasizes the vital role of blockchain technology in green finance, particularly in boosting energy efficiency and promoting renewable energy development [12]. The capabilities of blockchain, notably in ensuring transparency, accountability, and security, are instrumental in the efficient mobilization and allocation of green finance resources. This, in turn, aids in bolstering renewable energy projects, thereby contributing to greater energy efficiency and the advancement of sustainability initiatives.

Challenges in implementing blockchain in green finance include technological complexities, regulatory uncertainties, and concerns about blockchain's own environmental impact. There's also the challenge of integrating blockchain with existing financial systems and ensuring it aligns with green finance principles [13].

In conclusion, the integration of blockchain technology in green finance is revolutionizing the field, enhancing the effectiveness, transparency, and accountability of green finance initiatives and projects. The aforementioned studies provide valuable insights and evidence of the significant impact of blockchain technology in promoting sustainability, reducing carbon emissions, and enhancing renewable energy development, underscoring the technology's critical role in advancing green finance [7]–[12], [14].

### B. Regulatory Compliance and Supervision

Blockchain technology significantly advances regulatory compliance and supervision in various sectors by promoting transparency, accountability, and adherence to regulatory norms. Within the landscape of blockchain networks, the development of a credibility-based fuzzy comprehensive evaluation model enhances the selection process of oversight nodes, representing a significant contribution in this field. This innovation also entails designing a data block structure for efficient supervisor node replacement, thereby enhancing the effectiveness and reliability of systems like food safety supervision [15]. This approach represents a significant step in leveraging blockchain for complex regulatory environments.

The concept of embedded supervision in blockchain demonstrates its potential for automated compliance monitoring in tokenized markets. This system involves reading the ledger of the market, thus reducing the need for active data collection, verification, and delivery by firms [16]. Blockchain's core features, such as transparency and immutability, are leveraged to ensure adherence to regulations and enhance trust in financial systems.

In clinical research, blockchain's role extends to guaranteeing data authenticity and facilitating access for researchers and patients. The discussion includes regulatory considerations for blockchain in clinical research, emphasizing the need for compliant blockchain solutions [7]. Further, the possibility of automating procedures and information exchange among clinical trial participants through the integration of Ethereum smart contracts and IPFS is investigated, guaranteeing protocol compliance and data integrity [17].

The introduction of regulatory information retrieval (REG-IR) marks a significant advancement in ensuring legislative compliance, particularly within the context of EU/UK laws [18]. This approach leverages advanced BERT models, fine-tuned for specific in-domain classification tasks, to effectively navigate the complexities of continuously evolving legal frameworks. The application of these models in REG-IR demonstrates the considerable potential of blockchain technology in the field of regulatory technology (RegTech), offering innovative solutions for managing legislative changes.

Concurrently, studies conducted in the healthcare domain examine how blockchain applications conform to the General Data Protection Regulation (GDPR). This exploration emphasizes the imperative for blockchain developers to engage in informed design processes, ensuring that blockchain solutions are compliant with existing regulatory structures [19]. The study highlights the criticality of blockchain in augmenting interoperability and control over access to health data, all while maintaining the confidentiality and privacy of patient information. This dual focus on compliance and privacy underscores the multifaceted impact of blockchain in managing sensitive health data within stringent legal boundaries.

Key challenges include aligning blockchain technology with existing regulatory frameworks, addressing data privacy concerns, and ensuring compliance with standards like GDPR. There are also difficulties in developing effective supervision mechanisms for blockchain-based systems [20].

In conclusion, blockchain technology is proving to be a valuable tool in ensuring regulatory compliance and supervision across various domains, from food safety and finance to clinical research and healthcare. The technical advancements and innovative approaches discussed in the aforementioned studies underscore the potential of blockchain to revolutionize regulatory compliance, offering automated, transparent, and reliable solutions for meeting regulatory standards and enhancing trust in diverse sectors [7], [15]–[19], [21].

*C. Supply Chain Finance*

Blockchain technology is transforming the supply chain finance industry with cutting-edge approaches to enhancing transactional dependability and efficiency. A notable development in this field is the BCautoSCF platform, a blockchain-driven system designed for the auto retail industry. This platform exemplifies how blockchain can optimize workflows and enhance transaction security [22]. Additionally, there is an exploration of how blockchain is adopted in supply chain finance, with a particular focus on automating processes such as KYC, accounting, and transaction settlements, illustrating blockchain's pivotal role in addressing complex automation challenges within this realm [4].

A hybrid blockchain model combining PANDA and X-Alliance consensus algorithms is introduced, aimed at enhancing transaction processing, data reliability, and authority management in supply chain finance. This model is specifically designed for robust data ownership and change tracking [23]. Furthermore, the application of blockchain in the Beibu Gulf Region's supply chain finance is discussed, focusing on risk evaluation and financial game strategies [24].

Additionally, research shows the importance of an interdisciplinary approach to blockchain education, particularly in its application across various sectors including supply chain finance [25]. This perspective highlights the need for a comprehensive understanding of blockchain, addressing challenges and opportunities from diverse scientific angles.

A comprehensive literature review provides an in-depth analysis at the intersection of blockchain technology, supply chain finance, and legal frameworks. It addresses the operational, financial, and legal challenges in supply chain financing, exploring how blockchain can effectively mitigate these issues [26]. . Additionally, a detailed review of Distributed Ledger Technology (DLT) across various sectors, including supply chain finance, emphasizes its critical role in enhancing the reliability and availability of decentralized systems worldwide [27].

Challenges here include technical integration with existing supply chain systems, scalability issues, and ensuring data security and privacy. Adoption barriers also arise from lack of awareness, regulatory hurdles, and the diverse complexities of the blockchain technology [28].

In summary, the research collectively emphasizes how blockchain technology holds the potential to revolutionize supply chain finance. It accentuates a range of applications, from improving transaction security and efficiency to introducing innovative models and approaches for financial operations and education within the blockchain domain.

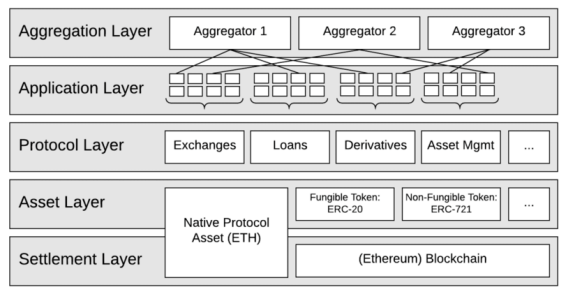

Fig. 2. Topological diagram of Blockchain and its applications

*D. Decentralized Finance (DeFi)*

The combination of blockchain technology and smart contracts is bringing about major changes in the banking industry,

especially within the Decentralized banking (DeFi) ecosystem. A detailed exploration of this ecosystem highlights the critical role of Ethereum's public smart contract platform which is illustrated in Fig 2 [29].Complementing this, a systematic analysis of the DeFi ecosystem is presented, outlining its foundational elements, operational protocols, and security dimensions, emphasizing the nuances of technical versus economic security [30].

In the intricate field of Decentralized Finance (DeFi), a deep dive into the blockchain oracle problem reveals its complexities and the quest for effective solutions. This analysis underscores the importance of standardization and the implementation of appropriate economic incentives to address inherent weaknesses [1]. Further enriching the discourse, a comprehensive examination of transaction reordering manipulations in DeFi is conducted. This includes categorizing and analyzing advanced mitigation schemes, offering insights into their diverse impacts on blockchain's integrity and functionality [31]. The research collectively contributes to a more robust understanding of DeFi's technical challenges and potential resolutions.

In the realm of Decentralized Finance (DeFi), a comprehensive introduction situates DeFi within the theoretical framework of permissionless blockchain technology. This includes providing a detailed taxonomical overview and identifying critical risk groups for stakeholders interested in decentralized financial applications [32]. Furthermore, a comprehensive comparison study of different Decentralised Autonomous Organisations (DAO) platforms on the Ethereum blockchain is carried out which highlights the distinguishing features and quantitative aspects of these platforms, providing valuable insights into their operational differences [33].

Another dimension of this discourse involves evaluating the role of financial intermediation in the context of DeFi. The analysis concludes that while DeFi transforms the landscape of financial intermediation, it does not entirely eliminate it. Instead, it reshapes the intermediation process, ensuring a more balanced distribution of power and preventing monopolistic control by any single entity. This nuanced understanding of DeFi's impact on traditional financial intermediation mechanisms offers a critical perspective on the evolving nature of financial systems in the blockchain era [34].

Decentralized Finance (DeFi) faces several key challenges including security risks from smart contract vulnerabilities, scalability issues amidst growing platform demands, and the need for regulatory compliance in an evolving financial landscape. Additionally, DeFi systems are susceptible to market manipulations, such as transaction reordering. Interoperability between different blockchain platforms, maintaining sufficient liquidity, user accessibility, and governance in decentralized autonomous organizations (DAOs) present further hurdles. The immutable nature of smart contracts also poses risks, as flaws in the code can have irreversible consequences, challenging the traditional roles of financial intermediaries in this new financial paradigm.

In conclusion, the thesis collectively offers a multifaceted insight into the world of DeFi, exploring its ecosystem, security, challenges, and the transformative role of blockchain and smart contracts. They contribute significantly to the understanding and advancement of blockchain technology in decentralized finance, underscoring its potential to revolutionize financial systems globally.

*E. IoT and Security*

A growing emphasis in the field of blockchain technology is on enhancing the efficacy and security of Internet of Things (IoT) devices. Studies in this domain investigate how blockchain can be applied in governance and various industries, including IoT, showcasing its capacity to address governance challenges and enhance services [35].The flexibility of blockchain is highlighted, extending its use beyond transactions to more robust governance frameworks.

Further contributions in this field include the development of a data security storage model for IoT systems, utilizing the architecture of the Fabric blockchain project [36]. This model is noted for its effectiveness and adaptability, offering improved data security for IoT systems. It addresses central challenges in IoT, such as transmission delays, single point of failure, and privacy concerns, enhancing the overall security and operational efficiency of IoT systems.

Enhancing blockchain's security aspects, one study introduces a novel approach for anomaly detection in blockchain networks, evaluating various ensemble learning methodologies [37]. This approach effectively combines multiple classifiers to improve anomaly detection, showcasing the use of advanced machine learning in strengthening blockchain security against threats and vulnerabilities.

Another research focuses on blockchain-based solutions for IoT security. It provides a comprehensive survey and categorization of prevalent security issues in IoT data privacy and the corresponding blockchain solutions [38]. This highlights blockchain's vital role in bolstering IoT security, protecting sensitive data, and reducing the risks of unauthorized access and data breaches.

Incorporating blockchain technology in IoT and security presents several challenges, including scalability issues due to the high volume of data generated by numerous IoT devices, and the complexity of integrating blockchain with diverse IoT ecosystems. Energy consumption is a significant concern, especially in proof-of-work blockchain systems, which may not be suitable for energy-sensitive IoT environments. Blockchain's inherent latency can impede real-time data processing in IoT applications, while privacy concerns emerge due to the transparent nature of blockchain transactions. Security risks exist, such as vulnerabilities in smart contracts and potential network intrusions. Regulatory and legal compliance is further complicated by the decentralised nature of blockchain, and compatibility issues may arise when combining different blockchain systems with Internet of Things protocols. Efficient data management on immutable blockchain ledgers and the associated operational costs further challenge the integration of blockchain in IoT and security domains.

In conclusion, this segment sheds light on the incorporation of blockchain technology to bolster security within the Internet of Things (IoT). From governance to data storage and anomaly detection, the application of blockchain stands as a pivotal element in reinforcing the security framework of IoT systems, ensuring their robustness, reliability, and resilience against potential threats and vulnerabilities.

*F. Blockchain in Emerging Markets*

Blockchain technology's influence in emerging markets extends across various sectors, notably in healthcare and the Metaverse. In healthcare, blockchain's role in enhancing global health security is crucial, offering improved data security and privacy. It facilitates efficient data sharing among stakeholders, while ensuring patient data anonymity, aiding in research and rapid detection of health hazards [39]. The capacity of the technology to maintain the anonymity of patient data is a critical advantage, assisting in healthcare research and the early detection and reporting of health hazards. This proactive strategy enables health authorities to take preventive steps as soon as possible, thereby strengthening the global health security system.

Blockchain technology is essential to removing major obstacles in the Metaverse, including resource requirements, application interoperability, and privacy and security issues [40]. MetaChain, a blockchain-based framework, is tailored to efficiently manage Metaverse interactions. Utilizing smart contract technology, it ensures seamless, secure transactions, thus enhancing user experience and facilitating the development of Metaverse applications. Importantly, the innovative sharding approach in the framework significantly improves blockchain scalability, tackling one of the technology's critical challenges.

Beyond 2020, blockchain technology holds promise for substantially enhancing climate markets through the integration of diverse national climate initiatives [41]. Research highlights blockchain's potential to support cost-effective and ambitious achievements in climate markets, promoting sustainable development and effective mitigation measures.

Within Software-Defined Networking (SDN), blockchain is becoming more widely known as a means of improving security in distributed network environments [42]. Discussions in this field focus on the framework of blockchain-based SDN, addressing its security challenges and future developments. The integration of blockchain in SDN strengthens network security, mitigates potential failure points, and ensures system robustness.

Blockchain's impact extends to the energy sector, notably in decentralized energy resource optimization in microgrid networks [43]. Use of blockchain architectures for peer-to-peer energy markets enabled decentralized coordination, improving resource scheduling and promoting sustainable energy use.

In Information Systems, blockchains have the to potential contribute to UN Sustainability Development Goals, highlighting adoption barriers and its transformative potential across industries [44]. Further, blockchain applications in medicine are investigated, focusing on data access, storage, security, and supply chain efficiency, underlining the importance of understanding blockchain's fundamental concepts and its future implications in medical care [45].

*G. Cross-Domain Applications*

In cross-domain applications, the versatility of blockchain technology is showcased through various innovative approaches. Leveraging consortium blockchain, a significant method is employed for lightweight authentication across cross-domain IoT environments, utilizing a digital token akin to cryptocurrency to generate trust and manage lifecycles [46]. This significantly bolsters security and operational efficiency in IoT systems, offering resilience against common threats.

Another significant contribution is a multifactor authentication protocol based on blockchain for the Industrial Internet of Things (IIoT), focusing on privacy preservation. This protocol uses multiple factors, converted into random numbers through hardware fingerprint encoding, to verify identities of IIoT devices securely and efficiently [47].

Furthermore, a model that prioritizes authentication across heterogeneous domains utilizes blockchain to bolster the integrity and robustness of the authentication process [48]. This model streamlines the procedure, making it more effective compared to traditional methods.

There is also a dynamic edge resource federation architecture that combines network slicing and blockchain for multi-domain IoT applications [49]. This innovative structure, named Hierarchical Integrated Federated Ledger (HIFL), ensures decentralized security and privacy in resource orchestration and service adjustment across multiple domains.

Furthermore, an exploration into Cross-Domain WiFi Sensing includes a comprehensive analysis of how various domains affect Channel State Information (CSI) [50]. It introduces a mathematical model for CSI, presenting a comprehensive framework for WiFi sensing systems across multiple domains to improve sensing accuracy in new environments.

It is proposed to merge a blockchain featuring permission controls with identity-based signatures and attribute-based access control within IoT access management systems. This integration involves establishing a dedicated blockchain ledger for each IoT domain, enhancing the capability of IoT devices to function as blockchain nodes and streamlining cross-domain access [51].

Finally, an introduction is made to a decentralized blockchain-oriented framework for permission assignment and access management within IoT [52]. This framework employs a hierarchical arrangement of local and global smart contracts, guaranteeing secure and effective cross-domain authentication and authorization in IoT environments.

III. METHODOLOGY

*A. Objective*

The principal goal of this research is to conduct a thorough analysis of the present-day implementations of blockchain technology in the financial domain. We aim to explore the various uses, benefits, and challenges associated with integrating blockchain into financial systems.

### B. Research Questions

How has blockchain technology been integrated into green finance? What role does blockchain play in regulatory compliance and supervision in finance? How is blockchain technology impacting supply chain finance? What are the hurdles and answers related to the adoption of blockchain in Decentralized Finance (DeFi)? How is blockchain technology being incorporated into financial applications involving Internet of Things (IoT) devices? What influence does blockchain technology have on emerging markets?

### C. Literature Search

To address the research questions, an extensive literature search was conducted using academic databases like PubMed, IEEE Xplore, and Google Scholar. The search utilized keywords such as "blockchain," "finance," "green finance," "regulatory compliance," "supply chain finance," "Decentralized Finance (DeFi)," "IoT," and "emerging markets."

### D. Selection Criteria

This research uses selected peer-reviewed articles published between 2010 and 2022, focusing on blockchain applications in finance. Emphasis is placed on research work that is highly cited, with a citation count threshold to ensure it's impact and relevance in the field.

### E. Data Extraction and Analysis

Data extraction involved analyzing selected papers and articles obtained through web scraping. The web scraper identified dominant themes by counting occurrences of specific keywords and their variations. The analysis categorized papers into themes like green finance, regulatory compliance, supply chain finance, DeFi, IoT integration, and impact in emerging markets. A comparative analysis, augmented by web-scraped data, identified common trends, challenges, and future directions.

### F. Limitations

This review is limited by the available peer-reviewed literature and the rapidly evolving nature of blockchain technology. The focus on highly cited papers may exclude newer research that has not yet gained widespread recognition, potentially affecting the findings or interpretations of the review.

### G. Discussion

The following subsections detail the applications, benefits, and obstacles of blockchain technology in the financial realm.

*1) Green Finance and Blockchain:* The utilization of blockchain in green finance highlights the technology's potential to revolutionize the sector. Blockchain's inherent transparency and immutability make it an ideal platform for tracking and verifying green investments, thereby reducing the risk of "greenwashing" [53]. This aligns with the growing demand for sustainable investment options and could significantly impact the mobilization of green finance.

*2) Regulatory Compliance and Blockchain:* Blockchain's potential in ensuring transparency and adherence to regulatory standards is evident. The hierarchical multi-domain blockchain network optimizes the election of supervision nodes, enhancing food safety supervision systems' efficiency and reliability [54].

Research finds the role of blockchain technology in enhancing anti-money laundering (AML) efforts within the banking sector. It focuses on how blockchain's inherent characteristics like transparency, immutability, and decentralized ledger can streamline AML processes. The research discusses the potential for real-time monitoring and analysis of financial transactions to detect and prevent fraudulent activities, thereby reinforcing regulatory compliance in banks [55].

Blockchain is being used in the healthcare research space to guarantee data validity and facilitate patient and researcher access. The regulatory aspects of blockchain in clinical research are covered in study, offering important insights into how to implement compliant blockchain solutions in this setting [56].Protocol compliance and data transparency are ensured by automating procedures and information sharing amongst clinical trial participants through the integration of Ethereum smart contracts and IPFS.

Research endeavors have delved into the conceptualization and implementation of an instantaneous payment framework underpinned by blockchain technology, with a pronounced emphasis on regulatory adherence. This inquiry scrutinizes the strategic deployment of blockchain's intrinsic attributes, namely decentralization, transparency, and cryptographic security, to engineer an expeditious and dependable payment infrastructure. Particular scrutiny is directed towards the harmonization of this system with prevailing regulatory imperatives, assuring scrupulous legal and fiscal compliance, while concurrently optimizing the efficacy of transactional processing mechanisms [57].

Our discussion also explores the compliance of blockchain applications in healthcare with the General Data Protection Regulation (GDPR), emphasizing the need for blockchain developers to make informed design choices to ensure compliance with existing regulatory frameworks [58]. This work emphasizes the pivotal importance of blockchain in improving interoperability and access control to health data, all while safeguarding patient privacy.

In conclusion, blockchain technology is proving to be a valuable tool in ensuring regulatory compliance and supervision across various domains, from food safety and finance to clinical research and healthcare. The technical advancements and innovative approaches discussed in the aforementioned studies underscore the potential of blockchain to revolutionize regulatory compliance, offering automated, transparent, and reliable solutions for meeting regulatory standards and enhancing trust in diverse sectors.

*3) Supply Chain Finance:* Blockchain technology significantly enhances supply chain finance by streamlining processes such as Know Your Customer (KYC) and transaction settlements. Its transparency ensures trustworthy trade among parties, reducing the risk of fraud. Smart contracts automate payment terms, leading to quicker settlements and improved working capital management. This technology offers real-time visibility into transactions, aiding in risk assessment and credit provision to small and medium enterprises (SMEs) [59].

*4) Decentralized Finance (DeFi):* DeFi platforms leverage blockchain to offer financial services without central intermediaries. They democratize finance by providing open access to lending, borrowing, and trading services. The programmability of smart contracts in DeFi allows for innovative financial products. However, challenges like the "oracle problem," where external data feeds interact with smart contracts, pose risks to data integrity and security in DeFi ecosystems [60].

*5) IoT and Blockchain:* The integration of IoT with blockchain technology in finance offers enhanced security and data integrity. IoT devices generate vast amounts of data, and blockchain provides a secure and immutable platform for storing this data. This integration enables transparent and tamper-proof record-keeping, essential for financial transactions. It also facilitates automated, real-time decision-making and improves the efficiency of financial services [61].

*6) Emerging Markets:* Blockchain technology in emerging markets presents opportunities for financial inclusion and economic development. An avenue for digital transactions that guarantees security and efficiency can be established in addition to enhancing access to financial services for the population. Blockchain's potential to reinforce supply chains, thereby ensuring transparency and trust in transactions, cannot be overlooked. However, it raises important considerations about its role in socio-economic development and potential risks [62].

The exploration undertaken in this discourse has delved into the multifaceted applications, advantages, and obstacles associated with blockchain technology within the financial realm. Through the resolution of these challenges and capitalization on the technology's merits, blockchain stands poised to transform the landscape of finance, promising the emergence of more transparent, efficient, and inclusive financial systems.

## IV. Addressing Gaps in Decentralized Trustless Insurance for Indian Agriculture

The precarious situation of Indian farmers, beset by the vagaries of droughts and floods leading to dire financial straits and, in extreme cases, suicides, presents a compelling case for innovative solutions. Despite governmental efforts, the persistence of these challenges underscores the need for a novel approach. The inherent vulnerabilities and inefficiencies associated with centralized, trust-based systems, including susceptibility to corruption and unethical practices, necessitate a paradigm shift. A decentralized, trustless mechanism emerges as a promising alternative, potentially revolutionizing agricultural insurance in India.

This research proposes the development of a blockchain-based insurance platform tailored for Indian farmers. Unlike conventional systems, this platform would be governed and executed by decentralized computing entities, eliminating the need for human intervention, centralized organizations, or databases. Leveraging a DeFi (Decentralized Finance) protocol, the platform operates through smart contracts on a decentralized blockchain network, ensuring immutability and execution across a global network of nodes. This architecture not only enhances transparency and security but also ensures data integrity through a consensus mechanism, effectively mitigating the risk of tampering.

The envisioned smart contract would enable farmers across India to contribute voluntarily, akin to insurance premiums, into a collective pool. Given the limitations of smart contracts in directly accessing external data, decentralized oracles play a crucial role in bridging this gap. These oracles, powered by a network of nodes, fetch and verify real-world information, such as weather conditions affecting agriculture, ensuring the system's integrity and responsiveness to actual events.

Furthermore, the adoption of a decentralized identity solution could safeguard against fraudulent data submission during the registration process, thereby enhancing the system's reliability. This comprehensive mechanism encompasses several stages, from farmer registration and premium contribution to the integration of real-time data via decentralized oracles, culminating in the automatic disbursement of claims based on predetermined criteria encoded in the smart contract.

This trustless model stands in stark contrast to traditional insurance schemes, eliminating the necessity for farmers to place their trust in intermediaries. The transparency and public verifiability of the smart contract code bolster confidence in the system's fairness and reliability. Additionally, the proposed technology stack, comprising Solidity for smart contract development, integration with Ethereum or compatible blockchains, and the utilization of React.JS for frontend development, provides a robust foundation for this decentralized application. While the incorporation of decentralized identity solutions is considered beneficial, it is not deemed indispensable for the initial deployment.

In summary, this initiative aims to fill a significant gap in the agricultural insurance landscape in India, leveraging blockchain technology to provide a secure, transparent, and equitable solution for farmers facing the challenges of climate-induced adversities. Through this decentralized approach, it is anticipated that the trust and participation of the farming community will be significantly enhanced, thereby contributing to their financial stability and resilience against natural calamities.

## V. Conclusion

The unfolding landscape of blockchain technology, as illustrated in this survey, demonstrates its far-reaching applications across diverse sectors and its transformative potential.

In green finance, blockchain offers increased transparency, mitigating the risk of greenwashing, and enables accurate tracking and verification of sustainable investments. Its integration with IoT not only fortifies security frameworks but also facilitates enhanced data collection and automation, with projects like Hyperledger Fabric pushing the frontier of IoT-blockchain integrations.

The realm of regulatory compliance is seeing a paradigm shift, with blockchain-driven models like hierarchical multi-domain blockchain networks streamlining processes, and smart contracts automating and ensuring adherence to standards. Supply chain finance has benefited from streamlined transactions, bolstered security, and optimized workflows, with initiatives like BCautoSCF exemplifying this transformation.

Decentralized Finance (DeFi) has emerged as a beacon of blockchain's potential to democratize and decentralize financial systems, with platforms like Ethereum underpinning many DeFi applications. Meanwhile, emerging markets are leveraging blockchain for diverse applications, from enhancing disease surveillance in healthcare to revolutionizing energy resource optimization in microgrid networks.

In cross-domain applications, the convergence of blockchain with other technologies has facilitated decentralized, secure, and efficient systems, from cross-domain IoT authentication models to advanced decentralized access control frameworks.

The integration of blockchain technology across various sectors, as explored in this survey, is not without its challenges. Scalability concerns, data interoperability issues, and regulatory complexities specific to each sector pose hurdles to blockchain's broader adoption. However, the ongoing advancements in consensus algorithms and cryptographic techniques, coupled with rising interdisciplinary collaborations, are paving the way for innovative and robust solutions. These developments are instrumental in overcoming current obstacles, further solidifying blockchain's role as a key driver in the next generation of technological advancements. The future landscape of blockchain, marked by its diverse applications and transformative potential, points towards a more decentralized, secure, and efficient technological ecosystem.

Despite these advancements, challenges like scalability, data interoperability, and sector-specific regulatory issues remain. Yet, the evolution of consensus algorithms, cryptographic enhancements, and increased interdisciplinary collaborations point towards robust solutions. In sum, blockchain's multifaceted applications, as highlighted in this review, indicate its pivotal role in shaping the next wave of technological innovations.

## References

[1] G. Caldarelli and J. Ellul, "The blockchain oracle problem in decentralized finance—a multivocal approach," *Applied Sciences*, vol. 11, no. 16, 2021.

[2] W. Baiod, J. Light, and A. Mahanti, "Blockchain technology and its applications across multiple domains: A survey," *Journal of International Technology and Information Management*, vol. 29, no. 4, pp. 78–119, 2021.

[3] G. Dorfleitner and D. Braun, "Fintech, digitalization and blockchain: possible applications for green finance," *The rise of green finance in Europe: opportunities and challenges for issuers, investors and marketplaces*, pp. 207–237, 2019.

[4] A. Rijanto, "Blockchain technology adoption in supply chain finance," *Journal of Theoretical and Applied Electronic Commerce Research*, vol. 16, no. 7, pp. 3078–3098, 2021.

[5] A. Iftekhar, X. Cui, Q. Tao, and C. Zheng, "Hyperledger fabric access control system for internet of things layer in blockchain-based applications," *Entropy*, vol. 23, no. 8, p. 1054, 2021.

[6] M. Campbell-Verduyn and F. Giumelli, "Enrolling into exclusion: African blockchain and decolonial ambitions in an evolving finance/security infrastructure," *Journal of Cultural Economy*, vol. 15, no. 4, pp. 524–543, 2022.

[7] W. Charles, N. Marler, L. Long, and S. Manion, "Blockchain compliance by design: Regulatory considerations for blockchain in clinical research," *Frontiers in Blockchain*, vol. 2, p. 18, 2019.

[8] L. Guo, S. Zhao, Y. Song, M. Tang, and H. Li, "Green finance, chemical fertilizer use and carbon emissions from agricultural production," *Agriculture*, vol. 12, no. 3, p. 313, 2022.

[9] Q. Xiong and D. Sun, "Influence analysis of green finance development impact on carbon emissions: an exploratory study based on fsqca," *Environmental Science and Pollution Research*, vol. 30, no. 22, pp. 61369–61380, 2023.

[10] Y. Fang and Z. Shao, "Whether green finance can effectively moderate the green technology innovation effect of heterogeneous environmental regulation," *International Journal of Environmental Research and Public Health*, vol. 19, no. 6, p. 3646, 2022.

[11] F. Mngumi, S. Shaorong, F. Shair, and M. Waqas, "Does green finance mitigate the effects of climate variability: role of renewable energy investment and infrastructure," *Environmental Science and Pollution Research*, vol. 29, no. 39, pp. 59287–59299, 2022.

[12] E. Rasoulinezhad and F. Taghizadeh-Hesary, "Role of green finance in improving energy efficiency and renewable energy development," *Energy Efficiency*, vol. 15, no. 2, p. 14, 2022.

[13] S. Bag, D. Viktorovich, A. Sahu, and D. A. Sahu, "Barriers to adoption of blockchain technology in green supply chain management," *Journal of Global Operations and Strategic Sourcing*, vol. ahead-of-print, 10 2020.

[14] M. J. Lahkani, S. Wang, M. Urbański, and M. Egorova, "Sustainable b2b e-commerce and blockchain-based supply chain finance," *Sustainability*, vol. 12, no. 10, p. 3968, 2020.

[15] Q. Tao, X. Cui, X. Huang, A. M. Leigh, and H. Gu, "Food safety supervision system based on hierarchical multi-domain blockchain network," *IEEE access*, vol. 7, pp. 51817–51826, 2019.

[16] R. Auer, "Embedded supervision: how to build regulation into blockchain finance," *Globalization and Monetary Policy Institute Working Paper*, no. 371, 2019.

[17] I. A. Omar, R. Jayaraman, K. Salah, M. C. E. Simsekler, I. Yaqoob, and S. Ellahham, "Ensuring protocol compliance and data transparency in clinical trials using blockchain smart contracts," *BMC Medical Research Methodology*, vol. 20, pp. 1–17, 2020.

[18] I. Chalkidis, M. Fergadiotis, N. Manginas, E. Katakalou, and P. Malakasiotis, "Regulatory compliance through doc2doc information retrieval: A case study in eu/uk legislation where text similarity has limitations," *arXiv preprint arXiv:2101.10726*, 2021.

[19] A. Hasselgren, P. K. Wan, M. Horn, K. Kralevska, D. Gligoroski, and A. Faxvaag, "Gdpr compliance for blockchain applications in healthcare," *arXiv preprint arXiv:2009.12913*, 2020.

[20] S. Han and S. Park, "A gap between blockchain and general data protection regulation: A systematic review," *IEEE Access*, vol. 10, pp. 103888–103905, 2022.

[21] D. Gozman, J. Liebenau, and T. Aste, "A case study of using blockchain technology in regulatory technology," *MIS Quarterly Executive*, vol. 19, no. 1, pp. 19–37, 2020.

[22] J. Chen, T. Cai, W. He, L. Chen, G. Zhao, W. Zou, and L. Guo, "A blockchain-driven supply chain finance application for auto retail industry," *Entropy*, vol. 22, no. 1, p. 95, 2020.

[23] J. Liu, L. Yan, and D. Wang, "A hybrid blockchain model for trusted data of supply chain finance," *Wireless personal communications*, pp. 1–25, 2021.

[24] R. Wang and Y. Wu, "Application of blockchain technology in supply chain finance of beibu gulf region," *Mathematical Problems in Engineering*, vol. 2021, pp. 1–10, 2021.

[25] B. Düdder, V. Fomin, T. Gürpinar, M. Henke, M. Iqbal, V. Janavičienė, R. Matulevičius, N. Straub, and H. Wu, "Interdisciplinary blockchain education: utilizing blockchain technology from various perspectives," *Frontiers in Blockchain*, vol. 3, p. 578022, 2021.

[26] I. Ioannou and G. Demirel, "Blockchain and supply chain finance: a critical literature review at the intersection of operations, finance and law," *Journal of Banking and Financial Technology*, vol. 6, no. 1, pp. 83–107, 2022.

[27] M. Gorbunova, P. Masek, M. Komarov, and A. Ometov, "Distributed ledger technology: State-of-the-art and current challenges," *Computer Science and Information Systems*, vol. 19, no. 1, pp. 65–85, 2022.

[28] K. Wannenwetsch, I. Ostermann, R. Priel, F. Gerschner, and A. Theissler, "Blockchain for supply chain management: A literature review and open challenges," *Procedia Computer Science*, vol. 225, pp. 1312–1321, 2023. 27th International Conference on Knowledge Based and Intelligent Information and Engineering Sytems (KES 2023).

[29] F. Schär, "Decentralized finance: On blockchain-and smart contract-based financial markets," *FRB of St. Louis Review*, 2021.

[30] S. Werner, D. Perez, L. Gudgeon, A. Klages-Mundt, D. Harz, and W. Knottenbelt, "Sok: Decentralized finance (defi)," in *Proceedings of the 4th ACM Conference on Advances in Financial Technologies*, pp. 30–46, 2022.

[31] L. Heimbach and R. Wattenhofer, "Sok: Preventing transaction reordering manipulations in decentralized finance," *arXiv preprint arXiv:2203.11520*, 2022.

[32] J. R. Jensen, V. von Wachter, and O. Ross, "An introduction to decentralized finance (defi)," *Complex Systems Informatics and Modeling Quarterly*, no. 26, pp. 46–54, 2021.

[33] Y. Faqir-Rhazoui, J. Arroyo, and S. Hassan, "A comparative analysis of the platforms for decentralized autonomous organizations in the ethereum blockchain," *Journal of Internet Services and Applications*, vol. 12, no. 1, pp. 1–20, 2021.

[34] L. Grassi, D. Lanfranchi, A. Faes, and F. M. Renga, "Do we still need financial intermediation? the case of decentralized finance–defi," *Qualitative Research in Accounting & Management*, vol. 19, no. 3, pp. 323–347, 2022.

[35] A. Razzaq, M. M. Khan, R. Talib, A. D. Butt, N. Hanif, S. Afzal, and M. R. Raouf, "Use of blockchain in governance: A systematic literature review," *International Journal of Advanced Computer Science and Applications*, vol. 10, no. 5, 2019.

[36] P. Wang and W. Susilo, "Data security storage model of the internet of things based on blockchain.," *Computer Systems Science & Engineering*, vol. 36, no. 1, 2021.

[37] S. Hisham, M. Makhtar, and A. A. Aziz, "Combining multiple classifiers using ensemble method for anomaly detection in blockchain networks: A comprehensive review," *International Journal of Advanced Computer Science and Applications*, vol. 13, no. 8, 2022.

[38] I. Al_Barazanchi, A. Murthy, A. A. Al Rababah, G. Khader, H. R. Abdulshaheed, H. T. Rauf, E. Daghighi, and Y. Niu, "Blockchain technology-based solutions for iot security," *Iraqi Journal for Computer Science and Mathematics*, vol. 3, no. 1, pp. 53–63, 2022.

[39] V. K. Chattu, A. Nanda, S. K. Chattu, S. M. Kadri, and A. W. Knight, "The emerging role of blockchain technology applications in routine disease surveillance systems to strengthen global health security," *Big Data and Cognitive Computing*, vol. 3, no. 2, p. 25, 2019.

[40] C. T. Nguyen, D. T. Hoang, D. N. Nguyen, and E. Dutkiewicz, "Metachain: A novel blockchain-based framework for metaverse applications," in *2022 IEEE 95th Vehicular Technology Conference:(VTC2022-Spring)*, pp. 1–5, IEEE, 2022.

[41] X. Dong, R. Mok, D. Tabassum, P. Guigon, E. Ferreira, C. Sinha, N. Prasad, J. Madden, T. Baumann, J. Libersky, *et al.*, "Blockchain and emerging digital technologies for enhancing post-2020 climate markets; world bank group: Washington, dc, usa, 2018."

[42] W. Li, W. Meng, Z. Liu, and M.-H. Au, "Towards blockchain-based software-defined networking: security challenges and solutions," *IEICE Transactions on Information and Systems*, vol. 103, no. 2, pp. 196–203, 2020.

[43] E. Münsing, J. Mather, and S. Moura, "Blockchains for decentralized optimization of energy resources in microgrid networks," in *2017 IEEE conference on control technology and applications (CCTA)*, pp. 2164–2171, IEEE, 2017.

[44] L. Hughes, Y. K. Dwivedi, S. K. Misra, N. P. Rana, V. Raghavan, and V. Akella, "Blockchain research, practice and policy: Applications, benefits, limitations, emerging research themes and research agenda," *International journal of information management*, vol. 49, pp. 114–129, 2019.

[45] A. Costa, "Emerging blockchain technology solutions for modern healthcare infrastructure," 2019.

[46] Y. Zhang, Y. Luo, X. Chen, F. Tong, Y. Xu, J. Tao, and G. Cheng, "A lightweight authentication scheme based on consortium blockchain for cross-domain iot," *Security and Communication Networks*, vol. 2022, pp. 1–15, 2022.

[47] Y. Zhang, B. Li, J. Wu, B. Liu, R. Chen, and J. Chang, "Efficient and privacy-preserving blockchain-based multifactor device authentication protocol for cross-domain iiot," *IEEE Internet of Things Journal*, vol. 9, no. 22, pp. 22501–22515, 2022.

[48] J. Liu, Y. Liu, Y. Lai, R. Li, S. Wu, and S. Mian, "Cross-heterogeneous domain authentication scheme based on blockchain," *Journal of Artificial Intelligence and Technology*, vol. 1, no. 2, pp. 92–100, 2021.

[49] R. Xu, Y. Chen, X. Li, and E. Blasch, "A secure dynamic edge resource federation architecture for cross-domain iot systems," in *2022 International Conference on Computer Communications and Networks (ICCCN)*, pp. 1–7, IEEE, 2022.

[50] C. Chen, G. Zhou, and Y. Lin, "Cross-domain wifi sensing with channel state information: A survey," *ACM Computing Surveys*, vol. 55, no. 11, pp. 1–37, 2023.

[51] S. Sun, R. Du, S. Chen, and W. Li, "Blockchain-based iot access control system: towards security, lightweight, and cross-domain," *IEEE Access*, vol. 9, pp. 36868–36878, 2021.

[52] G. Ali, N. Ahmad, Y. Cao, S. Khan, H. Cruickshank, E. A. Qazi, and A. Ali, "xdbauth: Blockchain based cross domain authentication and authorization framework for internet of things," *IEEE Access*, vol. 8, pp. 58800–58816, 2020.

[53] S. Li, R. Chen, Z. Li, and X. Chen, "Can blockchain help curb "greenwashing" in green finance? - based on tripartite evolutionary game theory," *Journal of Cleaner Production*, vol. 435, p. 140447, 2024.

[54] Q. Tao, X. Cui, X. Huang, A. M. Leigh, and H. Gu, "Food safety supervision system based on hierarchical multi-domain blockchain network," *IEEE Access*, vol. 7, pp. 51817–51826, 2019.

[55] A. Thommandru and D. B. Chakka, "Recalibrating the banking sector with blockchain technology for effective anti-money laundering compliances by banks," *Sustainable Futures*, vol. 5, p. 100107, 2023.

[56] W. Charles, N. Marler, L. Long, and S. Manion, "Blockchain compliance by design: Regulatory considerations for blockchain in clinical research," *Frontiers in Blockchain*, vol. 2, 2019.

[57] V. Zilnieks, "Concept of a regulatory compliant blockchain based instant payment system. limitations and compromises," in *2020 61st International Scientific Conference on Information Technology and Management Science of Riga Technical University (ITMS)*, pp. 1–7, 2020.

[58] L. Wang, Z. Guan, Z. Chen, and M. Hu, "Enabling integrity and compliance auditing in blockchain-based gdpr-compliant data management," *IEEE Internet of Things Journal*, vol. 10, no. 23, pp. 20955–20968, 2023.

[59] X. Chen and H. Liu, "Blockchain technology participates in the path and mode optimization of supply chain finance," *Journal of Innovation and Development*, 2023.

[60] A. Perdana, E. HU, and Rianto, "Decentralized finance (defi), strengths become weaknesses: a literature survey," *Jurnal RESTI (Rekayasa Sistem dan Teknologi Informasi)*, vol. 7, pp. 397–404, 03 2023.

[61] T. Saba, K. Haseeb, A. Rehman, and G. Jeon, "Blockchain-enabled intelligent iot protocol for high-performance and secured big financial data transaction," *IEEE Transactions on Computational Social Systems*, pp. 1–8, 2023.

[62] D. V. Chittipaka, S. Kumar, U. Sivarajah, J. Bowden, and M. Baral, "Blockchain technology for supply chains operating in emerging markets: an empirical examination of technology-organization-environment (toe) framework," *Annals of Operations Research*, vol. 327, pp. 1–28, 07 2022.